\documentstyle[pra,aps]{revtex}
\begin{document}
\draft
\title{Dimensional enhancement of kinetic energies}
\author{W. P. Schleich$^1$ and J. P. 
Dahl$^{1,2}$\footnote{Electronic address: jpd@kemi.dtu.dk}}
\address{
$^1$Abteilung f\"{u}r Quantenphysik, Universit\"{a}t Ulm,
D-89069 Ulm, Germany\\
$^2$Chemical Physics, Department of Chemistry, Technical University of
Denmark, DTU 207, DK-2800 Lyngby, Denmark}

\date{\today}
\maketitle

\begin{abstract}
Simple thermodynamics considers kinetic energy to be an extensive 
variable which is proportional to the number, $N$, of particles. 
We present a quantum state of $N$ non--interacting particles for 
which the kinetic energy increases quadratically with $N$. This 
enhancement effect is tied to the quantum centrifugal potential 
whose strength is quadratic in the number of dimensions of 
configuration space.
\end{abstract}



\section{Introduction}
 \label{intro}
The intensity of light radiated from independent dipoles is 
the number, $N$, of dipoles times the intensity of a single one. 
However, when the dipoles are located within a wavelength of the 
radiation the total intensity is $N^2$ times the intensity of a 
single dipole \cite{Dicke}. This superradiance effect is due to 
constructive interference between the individual dipoles. In the 
present article, we propose an enhancement effect of a similar 
strength for the kinetic energy of matter waves confined to a 
region whose radial extension in hyperspace is essentially 
independent of the number of particles.

According to simple thermodynamics, kinetic energy is an extensive 
variable, that is, in general the average kinetic energy of an 
ensemble of $N$ particles is linear in $N$ \cite{Tolman}. A 
similar dependence holds for the kinetic energy of an ordinary 
Bose-Einstein condensate (BEC) of $N$ particles. However, we now 
show that for a special quantum state of $N$ non-relativistic 
particles the kinetic energy increases as $N^2$. This quantum 
state is completely symmetric under exchange of the coordinates of 
the particles, but there is no interaction between the particles.

For the enhancement effect to occur, all particles need to have 
the same mass, but they do not necessarily have to be identical 
particles, Nevertheless, the state could also correspond to $N$ 
identical particles. In this case, it would be the state of $N$ 
bosons which are strongly entangled. 

Dimensional enhancement of kinetic energies is due to the wave 
nature of the atoms. It results from the form of the Laplacian in 
$D$ dimensions, giving rise to the quantum centrifugal potential. 
We illustrate this phenomenon using $N$ non-relativistic particles 
of identical mass, $M$, in three space dimensions. Here, we 
concentrate on the motional degrees of freedom, but do not take 
into account the internal structure of the particles. Hence, we 
deal with a $D=3N$ dimensional configuration space, and the wave 
function $\Psi = \Psi(x_1,x_2,\ldots,x_D)$ depends on $D$ 
coordinates.

The proposed effect is most conspicuous for $s$-states, that is, 
when the wave function depends on the hyperradius, $r = 
(x_1^2+x_2^2+\cdots+x_D^2)^{1/2}$, only. In this case, the wave 
function is completely symmetric under exchange of coordinates 
of the particles, corresponding to a bosonic state. We shall only 
consider $s$-states in the present work.

Our paper is organized as follows: In Sec.\ \ref{wf}, we lay the 
ground work for the calculation of the kinetic energy of an 
ensemble of particles, by introducing the concept of the radial 
wave function in hyperspace. We illustrate it by using two 
examples related to a BEC in an isotropic harmonic trap. In Sec.\ 
\ref{operator}, we then turn to the discussion of the operator of 
kinetic energy and cast it into a form which brings out the 
quantum centrifugal potential. The latter is proportional to the 
square of the number of dimensions. It is this potential that may 
cause the kinetic energy to be quadratic in the number of 
particles, as discussed in Section \ref{KE}.
 
In Sec.\ \ref{examples}, we evaluate the kinetic energy of three 
different radial wave functions. The two motivated by BEC, and 
denoted by $u_0$ and $u_1$, show a linear dependence on the number 
of particles. However, for the third radial wave function $u_2$, 
which is independent of dimensions, we find a kinetic energy that 
depends on the square of the number of particles. We dedicate 
Section \ref{enhance} to a discussion of the origin of this 
enhancement effect.

In Sec.\ \ref{dynamics}, we then turn to a discussion of the 
quantum dynamics, starting from the wave functions $u_0$, $u_1$, 
and $u_2$. We evaluate the time dependence of the average radial 
momentum following from these initial conditions. We note that the 
momentum corresponding to the BEC wave functions increases with a 
steepness that is proportional to the square root of the number of 
particles. In contrast, for the wave function $u_2$ we obtain the 
remarkable result, that the steepness depends quadratically on the 
number of particles. We conclude in Section \ref{summary} with a 
brief summary.


\section{Wave functions}
 \label{wf}
For an $s$-state, the motion of the $N$ particles is described by 
the normalized wave function $\Psi = \Psi(r)$ or, equivalently, by the 
radial wave function $u(r)$ defined by the relation
\begin{equation}
\Psi(r)= \frac{1}{\sqrt{S_D}}
\frac{u(r)}{r^{\frac{D-1}{2}}} \, ,
\label{wavefunction}
\end{equation}
where $S_D$ denotes the total solid angle in $D$ dimensions,
  \begin{equation}
S_D = \frac{2\pi^{D/2}}{\Gamma(D/2)}\, ,
 \label{SD}
  \end{equation}
and $u(r)$ is normalized such that 
  \begin{equation}
\int_0^\infty |u(r)|^2\,dr = 1.
  \end{equation}

We shall look look at three particular wave functions of the above 
type. The first one may be constructed from a normalized Gaussian 
wave function, 
  \begin{equation}
\varphi(x) = 
\left(\frac{\kappa^2}{\pi}\right)^{\frac14} 
e^{-\frac12 \kappa^2 x^2},
  \end{equation}  
in one dimension, by forming the product function 
  \begin{equation}
\Psi_0(r) = \varphi(x_1)\varphi(x_2) \cdots \varphi(x_D) 
= \left(\frac{\kappa^2}{\pi}\right)^{\frac{D}{4}} 
e^{-\frac12 \kappa^2 r^2},
  \end{equation}
with $r$ being the hyperradius. This wave function describes the 
state of a BEC of $N = D/3$ non-interacting particles in an 
isotropic magnetic trap at zero temperature \cite{trap}. The value 
of $\kappa$ is determined by the harmonic potential of the trap.

Our second $N$-particle wave function, $\Psi_1(r)$, may be constructed 
by retaining the wave function $\varphi(x)$ for $D-1$ space 
directions, while taking a wave function of the form $x^2\varphi(x)$ 
for the last direction. Symmetrization by forming the coherent sum 
$\sum_{i=1}^D x_i^2\exp(-\frac12 \kappa^2r^2)$, and subsequent 
normalization gives in fact 
  \begin{equation}
\Psi_1(r) = \frac{2\kappa^{(D+4)/2}}{\pi^{D/4}\sqrt{D(D+2)}} 
r^2 e^{-\frac12 \kappa^2 r^2}.
  \end{equation}

The radial wave functions associated with $\Psi_0(r)$ and $\Psi_1(r)$ 
are
\begin{mathletters}
  \begin{eqnarray}
u_0(r) & = & {\cal N}_0 r^{(D-1)/2} e^{-\frac12 \kappa^2 r^2}, 
 \label{u0} \\
u_1(r) & = & {\cal N}_1 r^{(D+3)/2} e^{-\frac12 \kappa^2 r^2},
\label{u1}
\end{eqnarray}
\end{mathletters}
with the normalization factors ${\cal N}_0$ and ${\cal N}_1$ being
  \begin{equation}
{\cal N}_0= \left[ \frac{2}{\Gamma\left(\frac{D}{2}\right)} 
\right]^\frac12\kappa^{\frac{D}{2}}, \quad 
{\cal N}_1= \left[ \frac{2}{\Gamma\left(\frac{D}{2} + 2\right)} 
\right]^\frac12\kappa^{\frac{D}{2}+2} .
  \end{equation}

We defer the presentation of the third $N$-particle wave function 
to Sec.\ \ref{IC}.


\section{Operator of kinetic energy}
 \label{operator}
We now consider the kinetic energies corresponding to the 
two above wave functions. The kinetic-energy operator,
\begin{equation}
\hat{T} = -\frac{\hbar^2}{2M}\Delta^{(D)},
\end{equation}
is determined by the Laplacian \cite{Sommerfeld}
\begin{equation}
\Delta^{(D)} = \frac{\partial^2}{\partial r^2} +
\frac{D-1}{r}\frac{\partial}{\partial r} +
\frac{\hat{\Lambda}^2}{r^2}\, ,
\label{Laplacian}
\end{equation}
in $D$ dimensions. The operator $\hat{\Lambda}^2$, which 
involves derivatives with respect to the $D-1$ angles of 
hyperspace, is proportional to the angular momentum operator in 
configuration space \cite{Avery}. 

As discussed in Ref.\ \cite{DS}, we may also write 
  \begin{equation}
\hat{T} = \frac{\hat{p}_r^2}{2M} 
+ V_Q(r) 
- \frac{\hbar^2}{2M}\frac{\hat{\Lambda}^2}{r^2} \, .
 \label{Tall}
  \end{equation}
Here, we have introduced the radial momentum
  \begin{equation}
\hat{p}_r = \frac12\left(\frac{\bbox{r}}{r}\cdot\hat{\bbox{p}} 
+ \hat{\bbox{p}}\cdot\frac{\bbox{r}}{r}\right) 
 = \frac{\hbar}{i} \frac{1}{r^{\frac{D-1}{2}}}
\frac{\partial}{\partial r} r^{\frac{D-1}{2}} ,
\label{pr}
  \end{equation}
and the {\em quantum fictitious potential} 
\begin{equation}
V_Q(r) = \frac{\hbar^2}{2M}\frac{(D-1)(D-3)}{4r^2}. 
\label{V_Q}
\end{equation}

We emphasize that the quantum fictitious potential also emerges 
directly from acting with the $D$-dimensional Laplacian of Eq.\ 
(\ref{Laplacian}) on the wave-function $\Psi(r)$ of Eq.\ 
(\ref{wavefunction}). This yields, in fact,
\begin{equation}
\Delta^{(D)}\Psi = \frac{1}{\sqrt{S_D}}\frac{1}{r^{\frac{D-1}{2}}}
\left[-\frac{\partial^2}{\partial r^2}
+\frac{(D-1)(D-3)}{4r^2}\right]u(r).
\label{laplace operator}
\end{equation}

Since $V_Q$ is proportional to the square of Planck's constant, it 
is a quantum potential with no classical analogue. Moreover, $V_Q$ 
depends inversely on the square of the hyperradius. This feature 
reminds us of the classical centrifugal potential, which gives 
rise to the non-inertial centrifugal force. Indeed, for $D\ge 4$ 
the potential $V_Q$ given by Eq.\ (\ref{V_Q}) is positive and thus 
corresponds to a repulsive force. This property suggests the name 
{\em quantum centrifugal potential}. In accordance with this, some 
authors \cite{Herschbach} absorb it into the last term of Eq.\ 
(\ref{Tall}). This, however, conceals the fact that it is a 
genuine part of the radial kinetic energy. 

We note that for $D=2$, the potential $V_Q$ is negative 
corresponding to an attractive force. This centripetal force is 
unique to two dimensions and counterintuitive to the classical 
notion of the centrifugal force always being repulsive. To capture 
this contradiction, we have coined the 
phrase {\em quantum anti-centrifugal potential} for the potential 
$V_Q$ in the case of $D=2$.
In Refs.\ \cite{Cirone} and \cite{Birula} we have focused on 
consequences of this attractive potential. However, in the present 
work we concentrate on the repulsive case corresponding to $D \ge 
4$. 

We recognize that one and three dimensions are also special: the 
potential $V_Q$ vanishes.


\section{Average kinetic energy}
 \label{KE}
The average kinetic energy
\begin{equation}
\langle\hat{T}\rangle = 
-\frac{\hbar^2}{2M}\int_0^\infty dr\ r^{D-1}
\int d\Omega_D\ \Psi^\ast \Delta^{(D)} \Psi
\label{T}
\end{equation}
of the particles described by $\Psi$ involves the Laplacian, Eq.\ 
(\ref{Laplacian}), and integrations over the hyperradius $r$ and 
the solid angle $\Omega_D$ in the $D$-dimensional hyperspace. With 
$\Psi = \Psi(r)$, as in the present work, the angular part of 
the Laplacian, that is, the operator $\hat{\Lambda}^2/r^2$ does not 
contribute to the integral. Hence, the kinetic energy may be said 
to be purely radial in hyperspace, and therefore also in the state 
spaces of the individual particles.

When we introduce the radial wave function $u(r)$, the 
average  kinetic energy defined by Eq.\ (\ref{T}) takes the form
\begin{equation}
\langle\hat{T}\rangle \equiv T = T_r + T_V .
\end{equation}
Here the contribution
\begin{equation}
T_r = \int_0^\infty dr\ u^\ast(r)\left[
-\frac{\hbar^2}{2M}\frac{d^2}{dr^2}\right] u(r)
\label{Tr}
 \label{para}
\end{equation}
is the average value of the operator $\hat{p}_r^2/2M$. We shall 
refer to it as the {\em para-radial kinetic energy}.

Furthermore, the contribution 
\begin{equation}
T_V = \int_0^\infty dr\ V_Q(r)|u(r)|^2
\label{Tv}
\end{equation}
results from $V_Q$.


\section{Examples}
 \label{examples}
The dimension $D$ of configuration space enters $V_Q$ 
quadratically. Hence, the total kinetic energy resulting from 
$T_V$ could in principle be quadratic in $D$. We recall that in 
the case of $N$ particles in three space dimensions, we deal with 
a $D=3N$ dimensional configuration space. Consequently, for 
$N\gg1$, the strength $S\equiv (D-1)(D-3)=(3N-1)(3N-3)\simeq 9N^2$ 
of the quantum non-inertial potential, is quadratic in the number 
of particles. According to Eq.\ (\ref{Tv}), the same thing may 
hold for the contribution $T_V$ to the kinetic energy.

\subsection{Thermodynamic case}
 \label{TC}
However, this feature strongly depends on the form and, in 
particular, on the $D$-dependence of the radial wave function $u$. 
For example, the wave functions $u_0$ and $u_1$ given by Eqs.\ 
(\ref{u0}) and (\ref{u1}), respectively, yield the radial kinetic 
energies 
\begin{equation}
T_r^{(0)}=\left[1 + \frac{1}{2(D-2)}\right]\epsilon,
\quad
T_r^{(1)}=\left[1 + \frac{1}{2(D+2)}\right]\epsilon ,
 \label{Tr1}
\end{equation}
and the quantum fictitious potential energies
\begin{mathletters}
\begin{eqnarray}
T_{V}^{(0)} & = & \left(\frac{D}{2}-1-\frac{1}{2}\frac{1}{D-2} 
\right)\epsilon, 
 \label{TV1a} \\ 
T_{V}^{(1)} & = & \left(\frac{D}{2}-3+\frac{15}{2}\frac{1}{D+2}
\right)\epsilon.
 \label{TV1b}
\end{eqnarray}
\end{mathletters}
Here, we have introduced the kinetic energy $\epsilon\equiv 
(\hbar\kappa)^2/(2M)$.

Hence, the total kinetic energies 
\begin{equation}
T^{(0)} = \frac{D}{2}\epsilon, \quad
T^{(1)} = \left(\frac{D}{2}-2+\frac{8}{D+2}\right)\epsilon
\label{TotalEn}
\end{equation}
are linear in $D$. In addition, the expression for $T^{(1)}$ involves 
correction terms independent of and inversely proportional to $D$.

In the limit of a large number of particles, that is 
$D=3N\gg 1$, we recover the thermodynamic result
\begin{equation}
T\simeq\frac{1}{2}D\epsilon=\frac{3}{2}N\epsilon 
 \label{T1}
\end{equation}
for both states.

It is interesting to note that the linear dependence on $D$ shown 
in Eq.\ (\ref{TotalEn}) is due to the contribution $T_V$, 
Eqs.\ (\ref{TV1a}) and (\ref{TV1b}), which emerges from the 
quantum centrifugal potential. The contribution from the 
para-radial kinetic energy is, according to Eq.\ (\ref{Tr1}), only 
weakly dependent on $D$.


\subsection{Enhancement case}
 \label{IC}
We now introduce the new wave function
\begin{equation}
u_2(r) = {\cal N}_2 \exp\left[-\frac12
\left(\frac{\beta}{r} +\kappa r\right)\right] ,
\label{u2}
\end{equation}
with the normalization constant
\begin{equation}
{\cal N}_2 = (\beta\kappa)^{-\frac{1}{4}}
\left(\frac{\kappa}{2 K_1(2\sqrt{\beta\kappa})}
\right)^{\frac{1}{2}}.
\end{equation}
Here, 
\begin{equation}
 K_n (\zeta) = \frac{1}{2} \int_0^{\infty} r^n
\exp\left[-\frac{\zeta}{2}\left(r+\frac{1}{r}\right)\right] 
\frac{dr}{r} 
\end{equation}
denotes the modified Bessel function \cite{Abram} of order $n$.

In contrast to $u_0(r)$ and $u_1(r)$, Eqs.\ (\ref{u0}) and 
(\ref{u1}), the wave function $u_2(r)$ is independent of the 
dimension $D$, that is, independent of the number of particles 
\cite{u2footnote}. As a consequence, the cancellation of one power 
of $D$ in the contribution $T_V$ to the kinetic energy which 
appears for $u_0$ and $u_1$, cannot take place for $u_2$. Indeed, 
when we substitute the wave function $u_2$ into the definitions 
Eqs.\ (\ref{Tr}) and (\ref{Tv}) of the energies $T_r$ and $T_V$, 
and perform the integrations, we arrive at
\begin{equation}
\label{Tr2}
T_r^{(2)}=\frac{1}{2\sqrt{\beta\kappa}}
\frac{K_2(2\sqrt{\beta\kappa})}{K_1(2\sqrt{\beta\kappa})}
\ \epsilon
\end{equation}
and
\begin{equation}
T_V^{(2)} = \frac{(D-1)(D-3)}{4\beta\kappa} \ \epsilon .
\end{equation}

The para-radial contribution $T_r^{(2)}$ is independent of $D$. In 
contrast, $T_V^{(2)}$ involves $D$, and hence the number $N$ of 
particles quadratically. This enhancement results from the quantum 
centrifugal potential and reflects the constraint, that as we 
squeeze more particles into the state we do not alter the radial 
wave function $u_2$. Indeed, $u_2$ is independent of $D$, and thus 
independent of $N$. Forcing additional particles into this state 
leads to a strong increase in energy.

This situation is, to some extent, analogous to the problem of 
confining charged particles to a given domain of space. Due to 
their Coulomb interaction, 
\begin{equation}
V_C(r)=\frac{q_1q_2}{r}\, ,
\label{coulomb}
\end{equation}
two charges $q_1$ and $q_2$ of the same polarity repel each 
other. This repulsion is proportional to the product of the 
charges and inversely proportional to their separation.

In the case of neutral particles, the quantum centrifugal 
potential of Eq.\ (\ref{V_Q}) is proportional to the product 
$(D-1)(D-3)$ of the dimensions. Hence, the dimension of 
configuration space plays the role of the charge. However, the 
analogy between $V_Q$ and $V_C$ breaks down in the dependence on 
the separation. Indeed, the hyperradius $r$ enters the denominator 
in a square, whereas in the Coulomb potential the separation 
enters only linearly.


\section{Origin of enhancement}
 \label{enhance}
This discussion suggests that we can interpret the energy 
$T_V^{(2)}$ associated with the quantum non-inertial potential as 
localization energy. The localization energy is defined as the 
energy necessary to localize a quantum particle, with the original 
energy $T_r^{(2)}$, in a domain of hyperspace whose radial 
extension is essentially independent of $D$. This energy obviously 
depends on the dimension of the space in which we want to localize 
the particle, because the volume of the unit sphere in $D$ 
dimensions decreases strongly for large $D$. Indeed, the 
expression (\ref{SD}) for $S_D$, together with an integration over 
$r$, easily gives the following expression for the volume 
  \begin{equation}
V_D = \frac{2\pi^{(D/2)}}{D\Gamma(D/2)} 
  \end{equation}
of the unit sphere. For $D>5$ this is a decreasing function of $D$.

It is interesting that in two dimensions $T_V^{(2)}$ becomes 
negative. Hence, we do not need to perform work in order to 
localize the particle in this case, but rather gain energy.

It is instructive to compare this quadratic dependence of the 
kinetic energy on the number of particles to the same scaling 
property of the total energy of $N$ fermions in a harmonic trap 
\cite{GW} of frequency $\Omega$. Indeed, when we assume that the 
fermions occupy the first $N$ states of the trap the total energy 
reads 
\begin{equation}
E=\hbar\Omega\sum_{j=0}^{N-1} \left( 
j+\frac{1}{2}\right)=\frac{\hbar\Omega}{2}
[N\left( N-1\right)+N] = \frac{N^2\hbar\Omega}{2}\, .
\label{Fermions}
\end{equation}

This simple calculation shows that the $N^2$-dependence arises 
from the fact that the particles are trapped in an external 
classical potential with a linear energy spectrum. In the case of 
the $N$ particles described by the wave function of Eq.\ 
(\ref{u2}), there is no classical potential present. Here, the 
$N^2$-effect results from the quantum centrifugal potential, and 
from the fact that the strength $S$ of this potential depends 
quadratically on the number of dimensions of configuration space.

Nevertheless, there exists a potential $V_2=V_2(r)$ 
for which the wave function $u_2$, Eq.\ (\ref{u2}), is an energy 
eigenstate. Indeed, when we differentiate $u_2$ twice with respect 
to $r$ we find that $u_2$ satisfies the time independent 
Schr\"odinger equation
\begin{equation}
\label{schroedeq}
\frac{d^2 u(r)}{d r^2}+\frac{2M}{\hbar^2}
\left[ E-V_2(r)\right]u(r)=0
\end{equation}
with energy $E=0$ and the potential
\begin{equation}
 V_2(r) = \frac{\hbar^2}{2M}\left[\frac{1}{4} 
\frac{\beta^2}{r^4}-\frac{\beta\kappa}{2r^2}-\frac{\beta}{r^3}+
       \left(\frac{\kappa}{2}\right)^2 \right].
\end{equation}
Since $u_2$ is independent of $D$, also the potential $V_2$ has to 
be independent of $D$.

We recall from Eq.\ (\ref{para}) that $d^2 u(r)/dr^2$ determines 
the para-radial kinetic energy rather than the total kinetic 
energy. To get the Schr\"odinger equation for the wave function 
$\Psi_2(r)$, as related to $u_2(r)$ through Eq.\ 
(\ref{wavefunction}), we draw on the expressions (\ref{Tall}) and 
(\ref{laplace operator}), and get:
  \begin{equation}
-\frac{\hbar^2}{2M}\Delta^{(D)}\Psi_2(r) 
+ \left[V_2(r)-V_Q(r)\right] \Psi_2(r) = E \Psi_2(r).
  \end{equation}
The confining potential in hyperspace is accordingly 
$V_2(r)-V_Q(r)$.


\section{Quantum dynamics}
 \label{dynamics}
So far, we have focused on quantum kinematics. Now we turn to a 
discussion of the dynamics resulting from the three radial wave 
function $u_0(r)$, $u_1$, and $u_2$, considered as initial wave 
functions when we switch off the confining potentials at 
time $t=0$, and hence allow for free-particle motion of the 
systems. Here we concentrate on the case of many particles, that 
is $D=3N\gg 3$.

The dynamics of this ensemble of free particles is governed by the 
Schr\"odinger equation
\begin{equation}
i\hbar\frac{\partial}{\partial t}\Psi(r,t)=
-\frac{\hbar^2}{2M}\Delta^{(D)}\Psi(r,t).
\end{equation}
By exploiting the relation (\ref{laplace operator}) for $\Psi$ we 
obtain the following expression for the time evolution of the radial 
wave function:
\begin{equation}
i\hbar\frac{\partial}{\partial t}u(r,t)=
\left[ -\frac{\hbar^2}{2M}\frac{\partial^2}{\partial r^2}+V_Q(r) 
\right]u(r,t).
 \label{TDS1}
\end{equation}

For short times, we may obtain a first approximation to $u(r,t)$ by 
replacing Eq.\ (\ref{TDS1}) with the equation
  \begin{equation}
i\hbar\frac{\partial}{\partial t}u(r,t)=
\left[W(r)+V_Q(r) \right]u(r,0), 
 \label{TDS2}
\end{equation}
where 
  \begin{equation}
W(r) = \frac{1}{u(r,0)}
\left[ -\frac{\hbar^2}{2M}\frac{\partial^2}{\partial r^2}
\right]u(r,0).
  \end{equation}
This quantity is similar to Bohm's quantum potential 
\cite{Holland}. Integration gives 
\begin{equation}
u(r,t) \simeq {\rm exp}\left\{ -\frac{i}{\hbar}[W(r)+ V_Q(r)]t 
\right\}u(r,0).
\label{u_rt}
 \label{urt}
\end{equation}

The formula (\ref{urt}) allows us to derive an analytical expression 
for the short-time behavior of the average radial momentum
\begin{equation}
\langle p_r \rangle (t)= \int_0^{\infty} dr\, 
r^{D-1}\int d\Omega_D \Psi^{*}(r,t)\hat p_r \Psi(r,t).
\end{equation}
By means of the definition Eq.\ (\ref{pr}) of the radial momentum 
$\hat p_r$ and the ansatz Eq.\ (\ref{wavefunction}) for the radial 
wave function, this expression reduces to
\begin{equation}
\langle p_r \rangle (t)= \int_0^{\infty} dr\, 
u^{*}(r,t)\left( \frac{\hbar}{i}\frac{\partial}{\partial r}\right) 
u(r,t).
 \label{prt}
\end{equation}
When we substitute the approximate solution Eq.\ (\ref{u_rt}) into 
this formula, we observe that $W(r)$ does not contribute to the 
integral because $u(r,0)$ is real valued. Hence, we arrive at
\begin{equation}
\langle p_r \rangle (t) \simeq
\int_0^{\infty} dr\, F_Q(r)|u(r,0)|^2t, 
 \label{prti}
\end{equation}
where 
  \begin{equation}
F_Q(r) = -\frac{dV_Q}{dr} 
= \frac{\hbar^2}{2M}\frac{(D-1)(D-3)}{2r^3}.
  \end{equation}
This quantity, determined by the derivative of the potential 
$V_Q$, is the quantum centrifugal force. The fact that the 
expression for $\langle p_r \rangle (t)$ is independent of $W(r)$ 
implies that we obtain the correct result for the short-time 
behavior of $\langle p_r \rangle (t)$ by neglecting the 
para-radial kinetic energy operator compared to the potential 
$V_Q$ in Eq.\ (\ref{TDS1}). The result of doing so is known as the 
Raman--Nath approximation \cite{Boss}. We also note that the 
integral preceding $t$ in Eq.\ (\ref{prti}) is nothing but 
$d\langle p_r \rangle (t)/dt$ evaluated at $t=0$. The expression 
(\ref{prti}) may therefore also be obtained by differentiating 
both sides of Eq.\ (\ref{prt}) with respect to $t$, while applying 
Ehrenfest's theorem \cite{Schiff}.

To sum up, we have found that the initial average radial momentum 
increases linearly in time. The slope of the increase is 
determined by the average centrifugal force of the initial state.

Let us now consider the specific wave functions $u_0$, $u_1$. and 
$u_2$. We start our discussion with the wave functions $u_0$ and 
$u_1$ of Eqs.\ (\ref{u0}) and (\ref{u1}), respectively. 
Substitution into the expression (\ref{prti}) for the average 
radial momentum yields
\begin{eqnarray}
\langle p_r \rangle^{(0)} (t) & \simeq & 
(D-1)\frac{\Gamma(\frac{D-1}{2})}{\Gamma(\frac{D}{2})}
\left( \frac{\epsilon t}{\hbar}\right)\hbar\kappa, 
\nonumber \\
\langle p_r \rangle^{(1)} (t) & \simeq & 
\frac{1}{2}(D-1)(D-3)
\frac{\Gamma(\frac{D+1}{2})}{\Gamma(\frac{D+4}{2})}
\left( \frac{\epsilon t}{\hbar}\right)\hbar\kappa.
\end{eqnarray}
With the help of the asymptotic formula \cite{Abram}
\begin{equation}
\Gamma(az+b) \sim \sqrt{2\pi} e^{-az}(az)^{az+b-1/2} ,
\end{equation}
we can evaluate the ratios 
\begin{equation}
\frac{\Gamma(\frac{D-1}{2})}{\Gamma(\frac{D}{2})} \sim 
\left(\frac{2}{D}\right)^{1/2}, 
\quad
\frac{\Gamma(\frac{D+1}{2})}{\Gamma(\frac{D+4}{2})} \sim 
\left(\frac{2}{D}\right)^{3/2}
\end{equation}
in the limit $D\gg 3$. For both wave functions, this yields 
\begin{equation}
\langle p_r \rangle (t) \sim \sqrt{2D}\left(\frac{\epsilon 
t}{\hbar}\right)\hbar\kappa.
\end{equation}
Hence, in the case of the initial wave function $u_0$ and $u_1$, 
the slope of the momentum increase is governed by the square root 
of the number of dimensions \cite{tlarge}.

Next, we turn to the wave function $u_2$, Eq.\ (\ref{u2}). In this 
case, the average radial momentum Eq.\ (\ref{prt}) takes the form
\begin{equation}
\langle p_r \rangle (t) \simeq 
\frac{(D-1)(D-3)}{2(\beta\kappa)^{3/2}}
\frac{K_2(2\sqrt{\beta\kappa})}{K_1(2\sqrt{\beta\kappa})}
\left(\frac{\epsilon t}{\hbar}\right) \hbar\kappa .
\end{equation}
Thus, the momentum corresponding to the wave function $u_2$ 
increases as $D^2$, and hence depends quadratically on the number 
of particles.

The explosion of the particles may be compared with the phenomenon of 
a Coulomb explosion \cite{Coulomb}. Atoms passing through a foil get 
stripped of some their electrons and become positively charged 
ions. Due to their Coulomb interaction, they repel each other and fly 
apart. In our case of neutral particles, it is the quantum centrifugal 
potential $V_Q$, Eq.\ (\ref{V_Q}), which causes the explosion.


\section{Summary}
 \label{summary}
We conclude by summarizing our main results. We have analyzed the 
kinetic energy and the dynamics of $N$ non-interacting particles 
in free space. We have found a quantum state for which the average 
kinetic energy increases quadratically with the number of 
particles. Moreover, this kinetic energy stored in the state gets 
transferred into outgoing radial momentum in the time evaluation 
pursuant to the preparation of the wave function. In case of $u_2$ 
the explosion is more violent than for $u_0$ and $u_1$, since the 
increase of momentum is proportional to $N^2$ rather than $N^{1/2}$ . 
Both effects---the dimensional enhancement of kinetic energies and 
the $N^2$-explosion---are consequences of the quantum centrifugal 
potential.


\section*{Acknowledgments}
We thank G. Alber, E. Arimondo, I. Bialynicki-Birula, F. Bopp,
J. Botero, A. Delgado, M. Mussinger, K. Vogel and W. Wonneberger 
for many fruitful discussions. J. P. D. gratefully acknowledges 
the support of the Alexander von Humboldt Stiftung.




\begin{thebibliography}{99}
\bibitem{Dicke}R. H. Dicke, Phys. Rev. {\bf 93}, 99 (1954).
\bibitem{Tolman}See, for example, R. C. Tolman, {\em Relativity, 
thermodynamics, and cosmology} (Clarendon, Oxford, 1934).
\bibitem{trap}See, for example, G. Baym and C. J Pethick, Phys.\ 
Rev.\ Lett.\ {\bf 76}, 6 (1996).
\bibitem{Sommerfeld}A. Sommerfeld, {\em Partielle
Differentialgleichungen der Physik} (Geest \& Portig, Leipzig,
1947); {\em Partial Differential Equations in Physics} (Academic
Press, New York, 1949).
\bibitem{Avery}J. Avery, {\em Hyperspherical Harmonics.
Applications in Quantum Theory} (Kluwer, London, 1989).
\bibitem{DS}J. P. Dahl and W. P. Schleich, Phys.\ Rev.\ A {\bf 
65}, 022109 (2002).
\bibitem{Herschbach}D. R. Herschbach, in {\em Dimensional Scaling 
in Chemical Physics}, p. 61, edited by D. R. Herschbach, J. 
Avery and O. Goscinski (Kluwer, Dortrecht, 1993). 
\bibitem{Cirone}M. A. Cirone, G. Metikas, and P. Schleich, Z. 
Naturf.\ {\bf 56a}, 48 (2001);
M. A. Cirone, J. P. Dahl, M. Fedorov, D. Greenberger and W. P.
Schleich, J. Phys.\ B. {\bf 35}, 191 (2002); 
M. A. Cirone, K. Rz\c{a}zweski, W. P. Schleich, F. Straub and J. A. 
Wheeler, Phys.\ Rev.\ A {\bf 65}, 022101 (2002).
\bibitem{Birula}I. Bia\l ynicki-Birula, M. A. Cirone, J. P. Dahl,
M. Fedorov, and W. P. Schleich, arXiv:quant-ph/0110116.
\bibitem{Abram}M. Abramowitz and I. Stegun, {\em Handbook of 
Mathematical Functions} (National Bureau of Standards, Washington, 
1964).
\bibitem{u2footnote}There are strict conditions on such a radial 
function. In particular, all its derivatives must vanish at $r=0$, 
in order that a proper $\Psi(r)$ of the form (\ref{wavefunction}) 
may be constructed for any $D$. The function $u_2(r)$ fulfills 
this requirement.
\bibitem{GW}See, for example, F. Gleisberg, W. Wonneberger, U. 
Schl\"oder, and C. Zimmermann, Phys.\ Rev.\ A {\bf 62}, 063602 
(2000).
\bibitem{Holland}See, for example, P. R. Holland, {\em The Quantum 
Theory of Motion} (Cambridge University Press, 1993).
\bibitem{Boss}W. P. Schleich, {\em Quantum optics in 
phase space} (Wiley-VCH, Weinheim, 2001).
\bibitem{Schiff}L. I. Schiff, {\em Quantum mechanics}, 3rd edition 
(Wiley, New York, 1968).
\bibitem{tlarge}For large values of $t$, the radial momentum settles 
at a constant value $p_\infty$. In the limit $D\gg 3$ this value may 
be determined from Eq. (\ref{T1}) by writing $T=p_\infty^2/2M$. This 
gives $p_\infty = \sqrt{D/2}\hbar\kappa$. Hence, we may also write 
$\langle p_r \rangle(t) \sim 2 p_\infty(\epsilon t/ \hbar)$. This 
is in accordance with the results of our discussion of $s$-wave 
implosion and explosion in ref.\ \cite{Birula}.
\bibitem{Coulomb}Z. Vager, R. Naaman, E. P. Kanter, Science {\bf 244}, 
426 (1989).
\end{thebibliography}
\end{document}